\documentclass[prodmode,acmtosn]{acmsmall} 

\usepackage{setspace}
\usepackage{hhline}
\usepackage{multirow}
\usepackage{multicol}
\usepackage{array}
\usepackage{xspace}   
\usepackage{graphicx} 
\usepackage[tight]{subfigure}
\usepackage{paralist} 
\usepackage{verbatim} 
\usepackage{url}      
\usepackage{wrapfig}  
\usepackage{booktabs}
\usepackage{colortbl}
\usepackage{tabularx}
\usepackage{amsmath}
\usepackage{color}
\usepackage{balance}
\usepackage{comment}
\usepackage[justification=centering,small]{caption}

\usepackage{setspace}
\usepackage{algorithm}
\usepackage{algpseudocode}
\usepackage{mathtools}

\acmVolume{9}
\acmNumber{4}
\acmArticle{1}
\acmYear{2015}
\acmMonth{3}

\begin{document}

\markboth{M. Hosseini et al.}{Joint Source Selection and Data Extrapolation in Social Sensing for Disaster Response}

\title{Joint Source Selection and Data Extrapolation in Social Sensing for Disaster Response}
\author{Mohammad Hosseini
\affil{University of Illinois at Urbana-Champaign, USA}
Nooreddin Nagibolhosseini
\affil{City University of New York, USA}
Amotz Barnoy 
\affil{City University of New York, USA}
Peter Terlecky
\affil{City University of New York, USA}
Hengchang Liu
\affil{University of Science and Technology, China}
Shaohan Hu
\affil{University of Illinois at Urbana-Champaign, USA}
Shiguang Wang
\affil{University of Illinois at Urbana-Champaign, USA}
Tanvir Amin
\affil{University of Illinois at Urbana-Champaign, USA}
Lu Su
\affil{University at Buffalo, USA}
Dong Wang
\affil{University of Notre Dame, USA}
Ramesh Govindan
\affil{University of Southern California, USA}
Raghu Ganti
\affil{IBM Research, USA}
Mudhakar Srivatsa
\affil{IBM Research, USA}
Charu Aggrawal
\affil{IBM Research, USA}
Tarek Abdelzaher
\affil{University of Illinois at Urbana-Champaign, USA}
Siyu Gu
\affil{Google}
Chenji Pan
\affil{LinkedIn}}


\begin{abstract}
This paper complements the large body of social sensing literature by developing means for augmenting sensing data with inference results that ``fill-in" missing pieces. It specifically explores the synergy between (i) inference techniques used for filling-in missing pieces and (ii) source selection techniques used to determine which pieces to retrieve in order to improve inference results. We focus on prediction in disaster scenarios, where disruptive trend changes occur. We first discuss our previous conference study that compared a set of prediction heuristics and developed a \textit{hybrid} prediction algorithm. We then enhance the prediction scheme by considering algorithms for sensor selection that improve inference quality. Our proposed source selection and extrapolation algorithms are tested using data collected during the New York City crisis in the aftermath of Hurricane Sandy in November 2012. The evaluation results show that consistently good predictions are achieved. The work is notable for addressing the bi-modal nature of damage propagation in complex systems subjected to stress, where periods of calm are interspersed with periods of severe change. It is novel in offering a new solution to the problem that jointly leverages source selection and extrapolation components thereby improving the results.
\end{abstract}

\category{C.2.2}{Computer-Communication Networks}{Network Protocols}

\terms{Design, Algorithms, Performance}

\keywords{Data extrapolation, wireless sensor networks, disaster response, social sensing, sensor selection}

\acmformat{Mohammad Hosseini, Nooreddin Naghibolhosseini, Amotz Barnoy, Peter Terlecky,
and Tarek F. Abdelzaher, 2015. Data Extrapolation in Social Sensing for Disaster Response.}

\begin{bottomstuff}
This work is supported by the Army Research Laboratory and was accomplished under Cooperative Agreement W911NF-09-2-0053, DTRA grant HDTRA1-10-1-0120, and NSF grant CNS 13-29886. The views and conclusions contained in this document are those of the authors and should not be interpreted as representing the official policies, either expressed or implied, of the Army Research Laboratory or the U.S. Government. The U.S. Government is authorized to reproduce and distribute reprints for Government purposes notwithstanding any copyright notation here on.

Author's addresses: M. Hosseini {and} T. F. Abdelzaher, Department of Computer Science,
University of Illinois at Urbana-Champaign; N. Naghibolhosseini  {and} A. Barnoy {and} P. Terlecky, Computer Science Department, City University of New York.
\end{bottomstuff}

\maketitle

\section{Introduction}
\label{sec:intro}

In this paper, we are addressing the problem of source selection and data extrapolation in participatory sensing applications, in the face of disruptive pattern changes, such as those that occur during natural disasters. We consider cases where resource limitations or accessibility constraints prevent attainment of full real-time coverage of the measured data space, hence calling for data extrapolation techniques to infer unknown variables from known ones. These techniques are complemented by source selection solutions that choose the subset of data to collect in order to maximize inference quality.

Many participatory sensing applications were investigated in recent years~\cite{Burke:Sensys,Huang:SenSys,Eisenman:SenSys,Mohan:SenSys,Ganti:MobiSys,Balan:MobiSys}. In participatory sensing, sources measure application-related state at locations of interest then usually report it at a later time (e.g., when they encounter a WiFi access point a few hours later). Hence, at any given time, the latest state of some points of interest may be unknown. Incomplete real-time coverage may also arise due to scarcity of sensing resources. For example, volunteers in a disaster-response application may survey and report locations of damage. If there are fewer volunteers than damage locations, the state of some of these locations will not be immediately reported. In such scenarios, one question is: can we infer the missing data?

Many time-series data extrapolation approaches are based on the assumption that past trends are predictive of future values. These approaches do not do well when disruptive changes occur. For example, a history of no traffic congestion on main highways of some city does not offer a good traffic predictor if a natural disaster causes a mass evacuation. An alternative recourse is to consider only spatial correlations. For example, certain city streets tend to get flooded together after heavy rain (e.g., because they are at the same low elevation), and certain blocks tend to run out of power together after a thunderstorm (e.g., because they share the same power lines). Understanding such correlations can thus help infer state at some locations from state at others when disruptive changes (such as a flood or a power outage) occur.

In our recent study~\cite{dcoss14}, we showed that system state in post-disaster scenarios alternates between periods of calm (when the past is a good predictor of the future) and periods of sudden change, as new parts of the infrastructure are damaged (e.g., due to aftershocks) or repaired. Hence, data extrapolation algorithms that rely predominantly on spatial correlations or predominantly on temporal correlations tend not to work consistently well, as the relative importance weights of temporal versus spatial correlations change significantly between periods of calm and periods of change. We argued  that such algorithms must switch intelligently between two extrapolation modes with different emphasis on temporal versus spatial correlations.

Of special interest is the case where correlations needed for extrapolation are themselves not known in advance, but are rather learned on the fly. The need for joint learning and extrapolation distinguishes this paper from some existing work~\cite{Aggarwal:DCOSS,Aggarwal:kdd,Biswas:IPSN,Kobayashi:Fuzzy} that predicts missing sensor values assuming a {\em previously known\/} correlation structure between sources, or a known temporal pattern. A new contribution lies in exploring the impact of sensor (source) selection algorithms on the efficacy of extrapolation schemes. We formulate source selection as an optimization problem, where the goal is to select the optimal subset of sources that allows us to best predict the state of all other sources. In this optimization problem formulation, we aim to minimize the total prediction error across all sources, subject to a retrieval budget constraint.

We apply the results to an example case study of a New York City crisis in the aftermath of Hurricane Sandy. Many gas stations, pharmacies, and grocery stores around NYC were closed after the hurricane, resulting in severe supply shortage that lasted several days. The outages were correlated, since different stores shared suppliers or power. Our study shows the degree to which synergistic data extrapolation and source selection could infer gas, food, and medical supply availability during the crisis in the absence of complete and fresh information. 


The remainder of this paper is organized as follows. We first present the general system model and illustrate prediction challenges in Section~\ref{sec:design}. Our \textit{hybrid} algorithm that addresses these challenges via appropriate switching between spatial and temporal extrapolation is presented in Section~\ref{sec:hybrid}. We then discuss the source selection problem for improving data prediction in Section~\ref{sec:selection1}, and propose two classes of source selection algorithms. Our evaluation results are presented in Section~\ref{sec:evaluation}. In Section~\ref{sec:soa}, we review related work. We conclude the paper in Section~\ref{sec:conclusion}.
\section{System Model}
\label{sec:design}
We consider a model of participatory sensing applications in which the reported state is binary. It is desired to obtain the state of several points of interest (POIs). A central collection node (e.g., the command center) collects the state from participants who make observations and report them later. The time when participants report their observations may vary. Measurements that are older than some threshold, are deemed stale. Hence, at any given time, there may be ``blind points'' in the POI map generated by participants, where fresh information is not available. The challenge is to infer the missing state automatically and accurately.
Two main difficulties characterize those scenarios~\cite{dcoss14}:

\begin{itemize}
\item {\em Disruptive change:\/} By definition, disasters are unique disruptive events that invalidate normal data trends, making prediction based on historical (time-series) trends largely incorrect.
\item {\em Scarcity of training data:\/} Since disasters are rare and generally unique, there is very little training data that one can rely on. Hence, the prediction algorithm (in the worst case) has training data available from the current disaster alone. This scarcity of data severely limits the complexity of prediction models that can be used.
\end{itemize}

We consider applications where today's information matters the most. For example, in the case of finding gas stations around NYC that are operational after hurricane Sandy, if one needed to fill up their car now, yesterday's gas availability would be of less use. The challenge is therefore to infer the {\em current\/} missing POI state.

We assume that old (and hence potentially stale) information on POI state is available. For example, in disaster response scenarios, volunteers might physically report back to the command center daily, which makes yesterday's information available at the center.
We call the maximum reporting latency, a {\em cycle\/}. Hence, by definition, the backend server knows the state of all POI sites in previous cycles, but has only partial information in the current cycle. This assumption simplifies our algorithmic treatment. It can easily be relaxed allowing for information gaps in previous cycles as well, since such gaps can always be filled in using the same extrapolation algorithm as the one applied to the current cycle.

\subsection{Problem Statement and Solution Challenges}
\label{sec:statement}
More formally, our participatory sensing system
can be characterized by a weighted graph $G = (\mathcal{T}, E)$, $|\mathcal{T}|=n$,
$|E|=m$, where the node set $\mathcal{T}$ represents the $n$ POIs. We assume
that set $\mathcal{T}$ is known and remains unchanged. The link set $E$ represents the correlations among POIs.

One way to compute links $E$, is to apply the Kendall's Tau statistical method \cite{Kendall:Biometrika}
to estimate correlations. More concretely, assume two POIs,
$x$ and $y$, have data $(x_1, x_2, \cdots, x_n)$ and $(y_1, y_2,
\cdots, y_n)$. The Kendall's
Tau correlation coefficient, denoted by $KT(x,y)$, can be
represented as:

\vspace{-0.5cm}
\begin{center}
\begin{equation}
KT(x,y) = 1 - \frac{1}{n}\sum_{i=1}^{n}XOR(x_i,y_i)
\label{equation:KT}
\end{equation}
\end{center}

Each edge $(x,y)$ between POI nodes $x$ and $y$ has a weight, $w_{xy} = KT(x,y)$, representing the correlation value. The link set $E$ may be reduced by setting a predefined threshold such that only links with correlations higher than the threshold are retained.

The extrapolation algorithm takes partial state of POI sites in the current cycle, historical data of POI sites in previous cycles, and the relationships (i.e., edges) learned so far as inputs. It then infers the current state of missing POI sites.

As argued above, scarcity of training data renders complex prediction models, such as ARIMA and various data mining 
models~\cite{arima13}, ineffective. For example, on the 4th day of a disaster, we have only 3 past training points, which might be fewer than the number of parameters in some models. This means that our prediction model would have to be very simple. Indeed a contribution of this work lies in arriving at a very simple model that works well with little data, as opposed to beating the current mature state of the art in time-series prediction from large data sets.

We first consider several obvious simple heuristics that can be used for extrapolation. To illustrate the impact of insufficient training data, we also consider ARIMA~\cite{arima13}, a standard (and powerful) time series analysis method for non-stationary processes, commonly used in complex forecasting tasks, such as forcasting financial systems~\cite{arima06}. The performance of these solutions will determine whether or not a new extrapolation approach is needed.

\begin{itemize}
\item
{\em Random}: It is the most trivial baseline in which the status of missing sites is guessed at random. It shows what happens when no intelligence is used in guessing.
\item
{\em BestProxy}: It uses the Kendall's Tau method to find actual pairwise (spatial) correlations between POIs and predicts missing state based on the state of the best neighbor (i.e., the POI that has the largest correlation with the one being predicted). It is an example of exploiting local spatial correlations, where state of an individual node is predicted from state of another (well-chosen) {\em individual node\/}.
\item
{\em Majority}: It computes the majority state of all known POIs and predicts all missing state to be the same as the majority state. This heuristic is another example of exploiting spatial correlations. It lies at the other end of the spectrum from {\em BestProxy}, in that it exploits a global notion of spatial correlations, where state of an individual node is predicted from {\em global state\/}.
\item
{\em LastKnownState}: It explores temporal correlations among POI sites. Namely, the predicted state today is set equal to the last known state.
\item
{\em ARIMA}: This, in principle, is one of the most general forecasting methods for time series data that assumes an underlying non-stationary process~\cite{arima13}.
\end{itemize}

\begin{figure*}[htb]
\centering
\begin{minipage}{0.32\textwidth}
\centering
\subfigure[][Distribution of gas outages ~~~~~~~~]{
\includegraphics[width=.9\columnwidth]{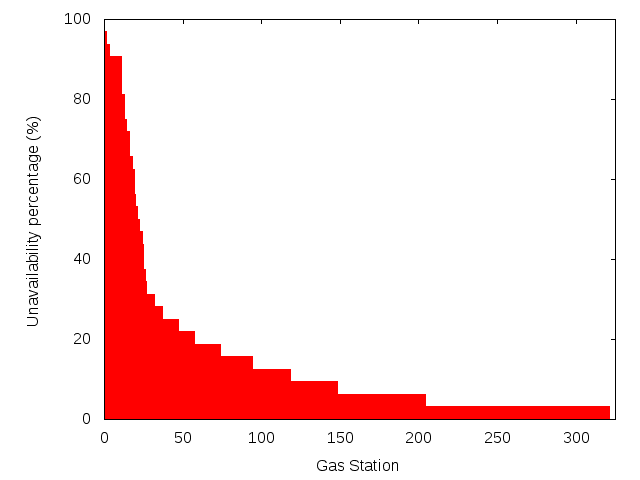}
\label{fig:fig1_gas}
} 
\end{minipage}
\begin{minipage}{0.32\textwidth}
\centering
\subfigure[][Distribution of food outages ~~~~~~~~]{
\includegraphics[width=.9\columnwidth]{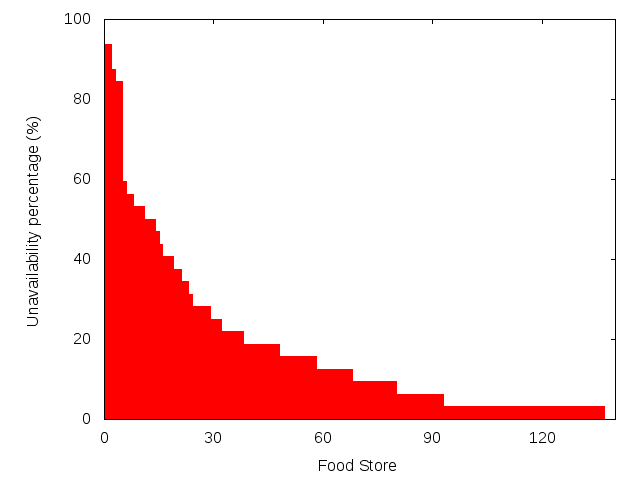}
\label{fig:fig1_food}
} 
\end{minipage}
\begin{minipage}{0.32\textwidth}
\centering
\subfigure[][Distribution of pharmacy outages]{
\includegraphics[width=.9\columnwidth]{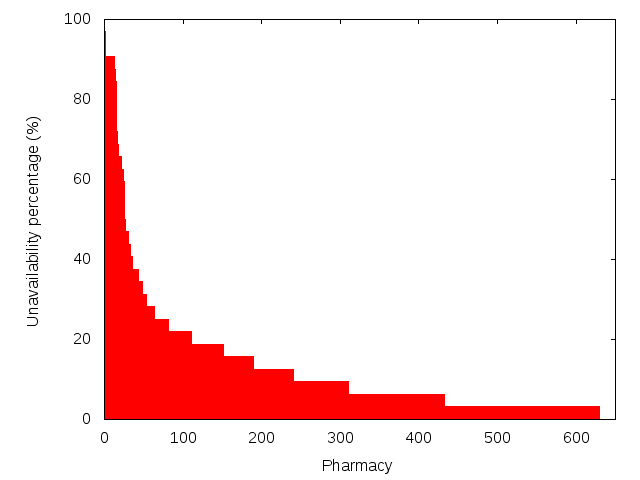}
\label{fig:fig1_pharma}
} 
\end{minipage}
\caption{Distribution of public services outages}
\end{figure*}

\noindent
Note that, we include {\em Random} to understand the baseline performance of a prediction algorithm that has no intelligence. {\em Best Proxy\/}, and {\em Majority\/} are different versions of algorithms that exploit spatial correlations. {\em LastKnownState} is a simple way of exploiting temporal correlations. {\em ARIMA} is a state of the art forecasting method. It is included to illustrate the inefficiency of such methods when training data is minimal. The performance of the above baselines is discussed next.

\subsection{Failure of Individual Baselines}

We evaluate our baselines through a real-world disaster response application. In November 2012~\cite{gasshortage:web}, Hurricane Sandy made landfall in New York City. It was the second-costliest hurricane in United States history (surpassed only by hurricane Katrina) and the deadliest in 2012. The hurricane caused wide-spread shortage of gas, food, and medical supplies as gas stations, pharmacies and (grocery) retail shops were forced to close. The shortage lasted about a month. Recovery efforts were interrupted by subsequent events, hence triggering alternating relapse and recovery patterns.

The daily availability of gas, food, and medical supplies was documented by the All Hazard Consortium
(AHC)~\cite{ahc:web}, which is a state-sanctioned non-profit organization focused on homeland security, emergency management, and business continuity issues in the mid-Atlantic and northeast regions of the United States. Data traces\footnote{Available at: http://www.ahcusa.org/hurricane-Sandy-assistance.htm} were collected in order to help identify
locations of fuel, food, hotels and pharmacies that may be open in specific geographic areas to support government and/or private sector planning and response activities. The data covered states including WV, VA, PA, NY, NJ,
MD, and DC. The information was updated daily (i.e., one observation per day for each gas station, pharmacy, or grocery shop). To give an example of the extent of damage, Figure~\ref{fig:fig1_gas} shows the distribution of the percentage of time that each of 300+ affected gas stations in the New York area was {\em unavailable\/} during the first {\em month\/} following the hurricane. We can see that 40 gas stations were not available for more than 1 week and some were out for almost the whole month. Similarly, Figure~\ref{fig:fig1_food} shows the distribution of outage for affected food stores and Figure~\ref{fig:fig1_pharma} shows the distribution of outage for affected pharmacies. 



With these POI sites and input data as ground truth, we evaluate the baselines described. The metrics we use are accuracy of inference and amount of data needed. 
We break time into cycles as discussed earlier. We set each cycle to a day to coincide with the AHC trace. We then plot the performance of the above baselines when a configurable amount of today's data is available (in addition to all historic data since the beginning of the hurricane).

\begin{figure*}[p]
\begin{minipage}{\textwidth}
\centering
\subfigure[][Error rate on Nov 3rd.]{
\includegraphics[width=0.482\textwidth]{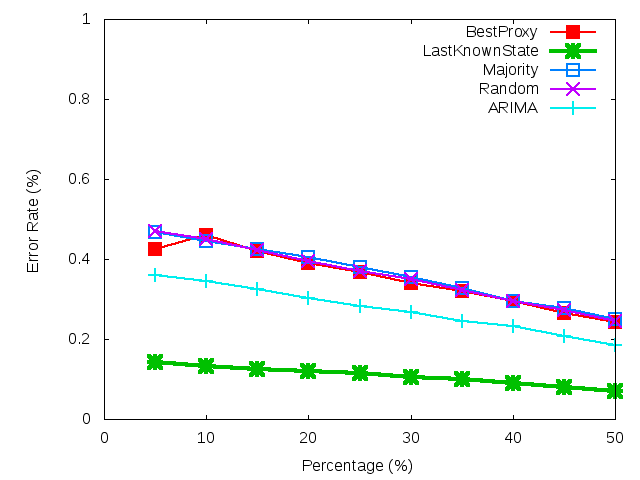}
\label{fig:gas4}
}
\subfigure[][Error rate on Nov 8th.]{
\includegraphics[width=0.482\textwidth]{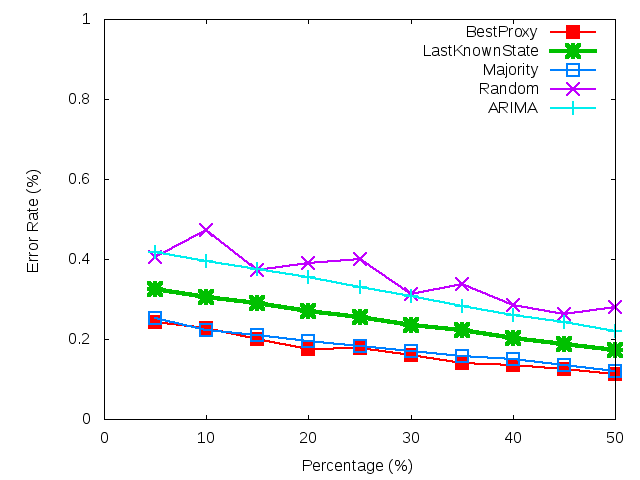}
\label{fig:gas9}
}
\caption{Comparing baselines to predict gas availability after Sandy}
\label{fig:accuracy_gas}
\end{minipage}
\begin{minipage}{\textwidth}
\centering
\subfigure[][Error rate on Nov 3rd]{
\includegraphics[width=0.482\textwidth]{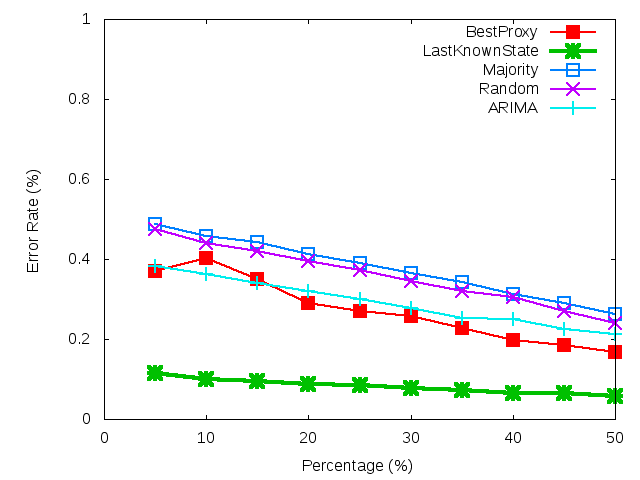}
\label{fig:food4}
}
\subfigure[][Error rate on Nov 8th]{
\includegraphics[width=0.482\textwidth]{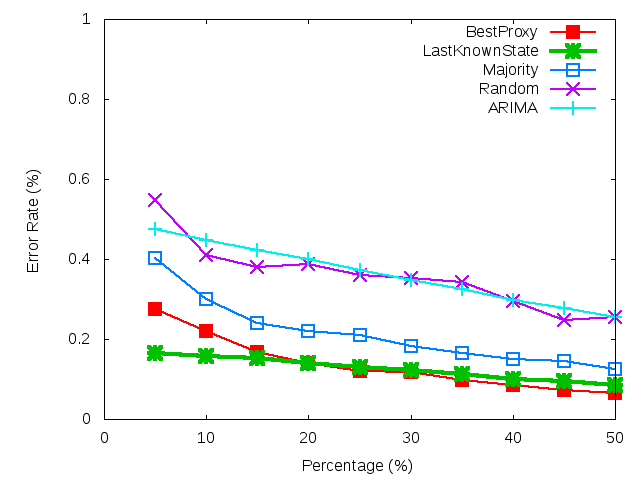}
\label{fig:food9}
}
\caption{Comparing baselines to predict food availability after Sandy}
\label{fig:accuracy_food}
\end{minipage}
\begin{minipage}{\textwidth}
\centering
\subfigure[][Error rate on Nov 3rd]{
\includegraphics[width=0.482\textwidth]{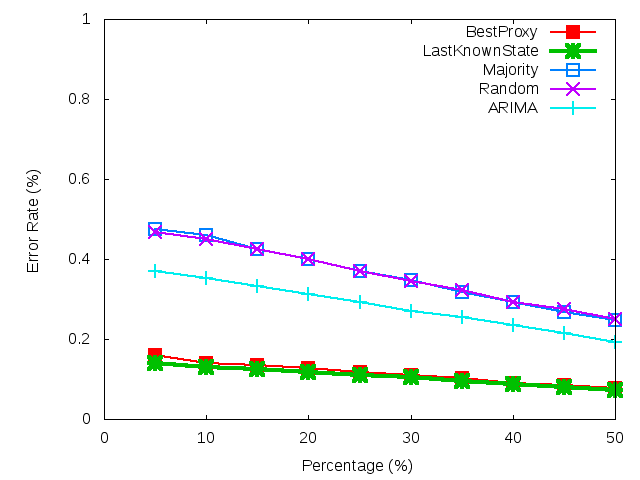}
\label{fig:pharma4}
}
\subfigure[][Error rate on Nov 8th]{
\includegraphics[width=0.482\textwidth]{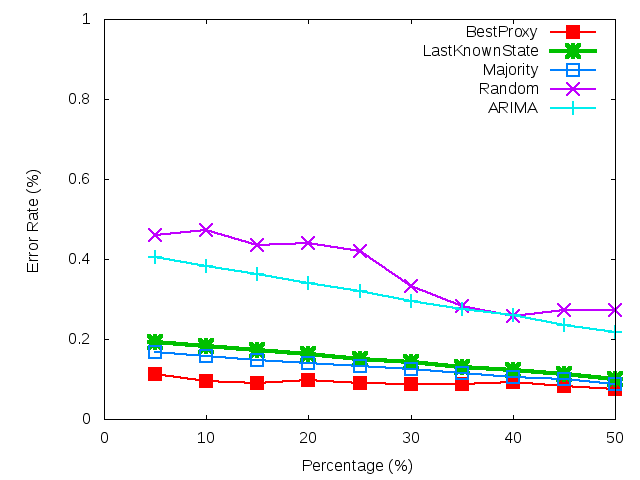}
\label{fig:pharma9}
}
\caption{Comparing baselines to predict pharmacy availability after Sandy}
\label{fig:accuracy_pharmacy}
\end{minipage}
\end{figure*}
We evaluate the solutions on November 3rd, and November 8th. November 8th corresponds to a period of disruptive change due to a second snow storm that hit after Sandy, causing massive temporary relapse of recovery efforts due to new power outages, followed by a quick state restoration to the previous recovery profile. November 3rd is an example of a period of little change, when damage was incurred but recovery efforts have not yet been effective. The same trend was observed for all datasets we have, namely, gas, pharmacy, and food.

Figure \ref{fig:accuracy_gas}, Figure \ref{fig:accuracy_food}, and \ref{fig:accuracy_pharmacy} plot the prediction error in availability of gas stations, food (grocery shops), and pharmacies, respectively. In each figure, sub-figures (a) and (b) refer to November 3rd and November 8th, respectively.
 
The reader is reminded that we assume that, on a given day, one knows the status of only a fraction of POIs (where the status refers to whether they are open or closed). The purpose is to extrapolate this data and find out the status of the remaining ones. The horizontal axis in the aforementioned figures varies the percentage of POIs whose status is known on the indicated day from 5\% to 50\%. To eliminate bias that may result from knowing the status of specific POIs, each point (corresponding to a specific percentage of POIs whose status is known) is an average of 50 different experiments. In each experiment, a different random set of POIs is selected as known (adding up to the required percentage). The results shown are the average of the 50 experiments.

Consider Figure~\ref{fig:accuracy_gas}-a and Figure~\ref{fig:accuracy_gas}-b, that illustrate the overall prediction error rate for {\em gas availability\/} on November 3rd and 8th, respectively, as a function of the percentage of POIs whose status is known that day. On the vertical axis, the performance of baselines is compared.

Figure~\ref{fig:accuracy_food}-a and Figure~\ref{fig:accuracy_food}-b similarly compare the performance of the baselines in predicting food availability on November 3rd and 8th. Figure~\ref{fig:accuracy_pharmacy}-a and Figure~\ref{fig:accuracy_pharmacy}-b compare the performance of the baselines in predicting pharmacy availability on November 3rd and 8th. 

It can be seen that no single baseline does consistently well in all figures. Specifically, LastKnownState does remarkably well on November 3rd, when the change was minimal from the day before. This is especially true for gas and food (grocery) availability prediction, where it beats the next heuristic by a wide margin. However, BestProxy does better on November 8th, when a second snow storm hits and its aftermath causes a lot of perturbation. BestProxy clearly outperforms LastKnownState that day for gas and pharmacy availability prediction, and ties for food availability prediction. Majority does poorly on November 3rd and better (but not best) on November 8th. Random does worse. Very interestingly, ARIMA does only marginally better than Random and much worse than the best heuristics on either day. This is attributed to the lack of sufficient training data, and the challenges caused by disruptive changes in the time-series.

The results confirm that algorithms that do spatial extrapolation (such as BestProxy) are better on days of more change, whereas algorithms that do temporal extrapolation (such as LastKnownState) are better on days of less change. The results also suggest that, due to lack of training data, complex prediction models that normally do well, such as ARIMA, are ineffective. We leverage these observations to guide the design of an algorithm that consistently offers the best performance. This algorithm appropriately adapts to periods of change versus periods of calm, and requires little training data. Note that, we do not aim to outperform any one heuristic at {\em all\/} times. Rather, our aim is to match consistently the best performing heuristic at any time, even though that heuristic changes, depending on circumstances. Such an algorithm is described next.

\section{A Hybrid Prediction Algorithm} 
\label{sec:hybrid}
The above study leads to two insights that help develop an algorithm for data extrapolation in disaster response scenarios:

\begin{itemize}
\item
{\em Insight \#1:\/} The first insight is that our algorithm should be able to switch between spatial and temporal prediction modes. On days with little change, LastKnownState does really well and should be the default prediction. On days where change is abundant, spatial correlations are more appropriate to use for prediction.
\item
{\em Insight \#2:\/}
The second insight lies in refining the notion of spatial correlations to be used for prediction. Since our default prediction is LastKnownState (i.e., no change), we need spatial correlations only to predict {\em change\/}. Hence, rather than using Kendall's Tau correlation to find a good proxy, we seek a proxy that helps predict change only. In other words, we seek a proxy whose {\em state changes\/} (and not overall state) are most correlated with those of the target to be predicted.
\end{itemize}  

\noindent
The second insight is intuitive in retrospect. Just because two gas stations were out of gas or out of power for a long time, does not mean their state changes are correlated. What's more indicative is whether or not they lost gas or power at the same time. The latter gives a better indication that if gas or power is restored to one, it may also be restored to the other.

More concretely, consider two POIs, $x$ and $y$, that have state $(x_1, x_2, ...,  x_n)$
and $(y_1, y_2, ..., y_n)$. Let $x_n$ be unknown (i.e., it has not yet been delivered). Let us define the change time series as $(dx_1, dx_2, ..., dx_n)$ and 
$(dy_1, dy_2, ..., dy_n)$, where $dx_i = x_i - x_{i-1}$ and $dy_i = y_i - y_{i-1}$ (we assume that $x_0 = 1$ and $y_0 = 1$ (everything was working before the disaster). To predict $x_n$ (or equivalently predict the change $dx_n$), we would like to find a proxy $y$, whose current status is known and whose changes are maximally correlated with changes in $x$. We can then use $dy_n$ to predict $dx_n$ and hence predict $x_n$. To do so, we compute $P (change~in~x | same~change~in~y)$ for all gas stations $y$ whose current state is known. This probability can be approximated by:
\begin{equation}
P (change~in~x | same~change~in~y) = \frac {count (dx_i = dy_i)}  {count (dy_i \neq 0)}
\label{eq:prob}
\end{equation}

\noindent
where $count()$ is a function that counts the number of times the condition in its argument was true for $1 \leq i \leq n-1$. The best proxy for (predicting change in) $x$ becomes the $y$ that maximizes the above probability. Let us call such a $y$, $y_{best}$. Let the resulting probability, $P (change~in~x | same~change~in~y^{best})$ be denoted $P^{best}$. Using insight \#1 above, the sought algorithm is as follows:

\begin{algorithm}[H]
\caption{\textsc{Enhanced Best Proxy} (x, n)}
\label{alg:BP2}
\begin{algorithmic}[1]
\State IF ( $P^{best} \geq$ threshold L )
\State use {\bf SpatialPrediction\/}
\State ELSE 
\State use LastKnownState (i.e., $x_n = x_{n-1}$)
\State
\State {\bf SpatialPrediction\/}
\State IF (($dy^{best}_n$ is not zero) AND ($y^{best}_{n-1} = x_{n-1}$))
\State THEN $x_n = y^{best}_n$
\State ELSE use LastKnownState (i.e., $x_n = x_{n-1}$)
\end{algorithmic}
\end{algorithm}

Lines 1 to 4 indicate that the algorithm alternates between spatial and temporal prediction depending on whether the best found proxy for the target $x$ is sufficiently good (i.e., better than a threshold, $L$). When spatial prediction is used, we predict that state of $x$ will change (i) if it was the same as the state of the best proxy, and (ii) if the state of that proxy changed. Otherwise, we predict no change.

It remains to derive the optimal value of the threshold, $L$. Let $M$ denote the fraction of POIs that had state = 1 in the last cycle. Hence, $1-M$ is the fraction of POIs with state = 0. Furthermore, let $F$ denote the fraction of POIs (that we are aware of so far) that change state in the current cycle. 
The optimal value of $L$ is one that minimizes misprediction probability.

The above algorithm mispredicts either (i) when spatial prediction is used and it is wrong, or (ii) when temporal (LastKnownState) prediction is used and it is wrong. Hence, misprediction probability, $P_m$, is equal to the sum of spatial misprediction probability, $P_{sm}$, and temporal misprediction probability, $P_{tm}$. Below, we compute these probabilities.
 
{\em Spatial Misprediction:\/} From line 7 of Algorithm 1, spatial misprediction occurs when (i) $P^{best}$ exceeds the threshold $L$ and (ii) the best proxy has the same state as $x$ in the last cycle, yet (iii) they have different states in the current cycle. Note that, the first two conditions are what invokes spatial prediction. The third condition causes that prediction to err.

Clearly, the probability of the first condition, $P (P^{best} > L)$, decreases with increasing threshold, $L$. Let us approximate $P (P^{best} > L) = 1 - L$. The probability of the second condition is simply $1 - 2M(1-M)$. Since $P^{best}$ is the probability of a correlated change in $x$ (given a change in the proxy), the probability of the third condition (a misprediction) is approximately $1 - P^{best}$. We know that $P^{best} > L$. Assuming that $P^{best}$ could be uniformly anywhere above $L$, we can replace $1 - P^{best}$ by $(1 - L)/2$. The spatial misprediction probability is then the product of probabilities of the three conditions above, leading to the expression:  
\begin{equation}
\label{eq:spatial-mis}
P_{sm} = (1 - L) [1 - 2M(1-M)] (1 - L)/2
\end{equation}

{\em Temporal misprediction\/} occurs when the algorithm resorts to temporal prediction and is wrong. According to the algorithm, temporal (LastKnownState) prediction occurs when (i) $P^{best}$ exceeds the threshold $L$, but (ii) the best proxy does not have the same state as $x$ in the last cycle, or when (iii) $P^{best}$ is less than the threshold $L$. In either case, a misprediction occurs if the state of $x$ changes (hence contradicting LastKnownState). The latter probability can be approximated by $F$, the fraction of nodes we know of that changed state today. Hence:
\begin{eqnarray}
\label{eq:temporal-mis}
P_{tm} & = & (1 - L) [ 2M(1-M)] F \\ \nonumber
       & + & [1 - (1 - L)] F
\end{eqnarray}
\noindent
Recall that misprediction probability, $P_m$, is the sum of $P_{sm}$ and $P_{tm}$. Hence, from Equation~(\ref{eq:spatial-mis}) and Equation~(\ref{eq:temporal-mis}), we get:
\begin{eqnarray}
\label{eq:mis}
P_m & = & (1 - L) [1 - 2M(1-M)] (1 - L)/2 \\ \nonumber
    & + & (1 - L) [ 2M(1-M)] F  \\ \nonumber
    & + & [1 - (1 - L)] F
\end{eqnarray}
\noindent
The optimal threshold, $L$, is one that minimizes the above probability. It can be found by setting the derivative of the above function to zero and enforcing the natural constraints on values of probability (that they are between 0 and 1). In other words:
\begin{eqnarray}
\label{eq:d-mis}
\frac {d P_m} {d L} & = & - (1 - L)[1 - 2M(1-M)] \\ \nonumber
    & - & [2M(1-M)] F  \\ \nonumber
    & + & F = 0
\end{eqnarray}
\noindent
subject to the constraint $0 \leq L \leq 1$. After some rearranging and algebraic manipulation, we get:
\begin{equation}
\label{eq:T}
L = 1 - F
\end{equation}
\noindent 
Unfortunately, we do not know the probability of change, F, in advance. In the absence of further knowledge, we can design for $F = 0.5$. In this case, L = 0.5. We hold off on evaluating this algorithm until we describe the next component of our scheme; namely, source selection. The source selection problem addresses the following question: what if we could choose which 5\% or 50\% of POIs to retreive today's information from. Which subset should we choose in order to maximize accuracy of prediction of missing POI state? The answer is investigated below.

\section{The Source Selection Problem}
\label{sec:selection1}
In the previous sections, we assumed that, on a given day, one knows the status of only a fraction of POIs. Hence, a \textit{random} set of POIs was selected as known, adding up to the required percentage (referring to the horizontal axis in Figures~\ref{fig:accuracy_gas} to~\ref{fig:accuracy_pharmacy}). This is consistent with cases where we have no control over the subset of POIs whose state we know. In some cases, however, we do have control. For example, we might be able to fly a drone to selected destinations to take pictures of their state. This has a cost, and we may have a total cost budget (e.g., limited time or money) that we are willing to spend on this search. Hence, the problem becomes one of optimizing the set of POIs whose state is retrieved such that the accuracy of prediction of the remaining POIs is maximized. In this section, we discuss this source selection problem. Similar to our initial setting, our participatory sensing system is characterized by a weighted directed graph $G = (\mathcal{T}, E)$ representing a network of $|\mathcal{T}|= n$ sources (i.e., POIs) and $|E|=m$ logical links between their states. Thus, each vertex $\tau_i \in \mathcal{T}$ represents a POI. Each edge $e_{ij} = (\tau_i, \tau_j) \in E$ represents the correlation between states of two POIs, $\tau_i$ and $\tau_j$. We omit links between POIs whose correlation is below a threshold. 

In our generalized source selection setup, each node $\tau_i$ is associated with a selection cost $c_i$, and an importance value of $v_i$, and each edge $e_{ij}$ is associated with a correlation $p_{ij}$, when using $\tau_i$ to predict $\tau_j$. We note that $c_i$ and $v_i$ are set in an application-specific manner. The cost $c_i$ may reflect the difficulty in querying the POI. The importance value $v_i$ reflects the importance of the POI. The problem of selecting the optimal set of POIs to query in order to make predictions about all others for which the current state is unknown can be formulated as an optimization problem, where we minimize the total prediction error of the whole system by selectively choosing a particular subset of POIs, $S \subseteq \mathcal{T}$, whose current states are retrieved such that total cost does not exceed some budget constraint, $W$.

In order to apply our optimization algorithms, it is necessary to model the prediction coefficient, or correlation on the links between POI states in this network. As previously mentioned, the prediction coefficient on the link from $\tau_i$ to $\tau_j$ is denoted by $p_{ij}$. We use both the Kendall's Tau and Hybrid correlation methods between the time series of both POIs to compute the value of $p_{ij}$. Under this definition, similar to our initial settings, assume two POIs $\tau_i$ and $\tau_j$ have time series $(x_1,…,x_t)$ and $(y_1,…,y_t)$ as the histories learned so far, respectively. Each edge ($\tau_i$, $\tau_j$) has a weight $p_{ij} = KT (\tau_i, \tau_j)$ or $p_{ij} = H (\tau_i, \tau_j)$ representing the Kendall's Tau or our enhanced Hybrid correlation coefficients. Once the graph has been created and the corresponding correlations have been calculated, they are later used by our source selection algorithms to estimate the state of the unknown POIs.

Our source selection scheme is a generalization of the knapsack problem (when the prediction edges are zero) or the dominating set problem (when the prediction edges are zero and one). Both of these problems are NP-hard but efficient approximation algorithms can be utilized, so this approach is computationally feasible.

Different variations of the knapsack problem has been applied to certain contexts. Y. Song \textit{et al.} in their paper \cite{4753629} investigated the multiple knapsack problem and its applications in cognitive radio networks. Hosseini \textit{et al.} \cite{mmsys13,movid14,mmve15} applied the concept of multiple-choice knapsack problem to the context of multimedia streaming applications to save bandwidth and energy.
X. Xie and J. Liu in their paper \cite{4223174} studied QKP, while also in \cite{6848188}, the authors applied the concept of m-dimensional knapsack problem to the packet-level scheduling problem for a network, and proposed an approximation algorithm for that. We build on top of these work, and develop multiple approximation algorithms which we describe in this section.

Let $S$ be the set of all selected sources, and $W$ be the budget. For each source $\tau_i$, we calculate a weighted value $v'_i$. This is the contribution that selecting $\tau_i$ alone would make to the predictability of the whole system if it is chosen. We define two general heuristics, namely \textit{Static}, and \textit{Dynamic} greedy algorithms, to calculate $v'_i$.

For the \textit{Static} greedy heuristic, we calculate $v'_i$ for each node as the sum of its importance value, $v_i$, and the importance value that this node can predict from the other nodes. To find the predicted value, the algorithm runs on all the nodes and gives a portion of value of each node to its maximum predictor given on the prediction edge (i.e. the neighbor node with maximum incoming edge weight). The idea here is that the error of predicting POI $\tau_i$ from a POI $\tau_j$ is magnified somehow by the importance value $v_i$ of POI $\tau_i$. We iterate over all nodes so finally $v'_i$ for all the nodes will be calculated with the following formula:

\begin{equation}\label{Prediction Values Under Static Greedy}
\begin{split}
& (i,j) \in E \iff\ There\ is\ an\ edge\ from\ \tau_i\ to\ \tau_j \\
& \tau_i \in S \iff \tau_i\ is\ selected \\
& p_{ij}\gets Percentage\ of\ prediction\ of\ node\ \tau_j\ if\ \tau_i\ is\ selected \\
& v'_{i}=v_{i}+\sum_{(i,j) \in E}(p_{ij}v_{j})
\end{split}
\end{equation}
Which is then used to select nodes given their ratio of ${v'_i}/{c_i}$.

The \textit{Dynamic} greedy heuristic is similar to the Static version, with the only difference being that in the dynamic version, after each round of selecting the best source candidate, $v'_i$ is updated for all other sources. So the first step is similar to the static greedy algorithm in that in the beginning, the algorithm calculates $v'_i$ for all sources. Unlike the static version which continuously selects the best source candidate from the set of unselected source at each round, the Dynamic greedy algorithm updates $v'_i$ for all unselected sources after the first selection. It then selects the second best candidate, updates $v'_i$ for all others in the set of unselected sources, and repeats this process until the budget is satisfied.

To update the value of the unselected sources at each round of this iteration, the algorithm checks all neighbors of each unselected source. Assume $\tau_h$ is the current unselected source which is being considered. Then, if at least one of $\tau_h$'s neighbors has been selected, the algorithm deducts the portion of the highest prediction of its neighbors from $\tau_h$. In the next step, the algorithm calculates $v'_h$ for all the unselected sources $\tau_h$. Starting from the first unselected source, it adds the portion of the value of each unselected source to its maximum predictor, and continues for all sources. If the maximum predictor is already selected, it is ignored, thus the next unselected source will be considered. If some pre-selected sources have already predicted this source but the maximum predictor is another source, then the difference of the maximum prediction value and prediction value of the maximum predictor of this source is added. Otherwise, the portion of the prediction value is added to the maximum predictor. The mathematical formula for the prediction value of each unselected source in each round is the following:

\begin{equation}\label{Prediction Values Under Dynamic Greedy}
\begin{split}
& p'_{i} = \max~_{(j,i) \in E\ \wedge\ \tau_j \in S}\ (p_{ji})\ or\ 0\\
& v'_{i}=(1-p'_{i}).v_{i}+\sum_{(i,j) \in E}(\max(p_{ij}-p'_{j},0).v_{j})
\end{split}
\end{equation}

To determine the optimal set of sources, this algorithm sorts the list of sources by $v'_i/c_i$ from largest to smallest. For ease of notation in the following, suppose that the sources are re-indexed so that the sorted list is $\tau_1,\tau_2,\ldots,\tau_n$. If $c_{\tau_1}\leq W_0= W$ then there is enough unused budget to select $\tau_1$, so the selected source $x_1$ has $c_{x_1}=c_{\tau_1}$ and contributes $v'_1$ to the average prediction value of the whole system. This leaves an unused budget of $W_1=W_0-c_{\tau_1}$ for the remaining sources after $x_1$. The algorithm repeats this process for $\tau_2, \tau_3,\ldots$ until some sources $\tau_{\ell}$ cannot be selected within the remaining budget $W_{\ell-1}$. Finally the total value and other statistics associated with the selected set of sources are calculated. 

Algorithm \ref{static} and Algorithm \ref{dynamic} is the psuedocode for Static and Dynamic greedy, respectively.

\begin{algorithm}[t]
 \caption{Static Greedy}
 \label{static}
\begin{algorithmic}[1]
\State $\mathcal{T}$: list of all sources $\tau_i$
\State $C$: list of covered sources
\State $S$: set of selected sources
\State $N_i$: set of neighbours of source $\tau_i$
\State $v_i$: importance value of source $\tau_i$
\State $c_i$: cost of source $\tau_i$ 
\State $e_{ij}$: prediction edge (using source $\tau_i$ to predict source $\tau_j$)
\\
\For {each $\tau_i \in \mathcal{T}$}
	\State $v'_x \gets v_x + e_{xi} v_i$ such that $e_{xi}= \max (e_{ji})~\forall \tau_j \in N_i~or~0$
\EndFor
\State $cost \gets 0$
\While {$|\mathcal{T}| \neq |C|$}
	\State $x \gets j$ such that $(v'_j/c_i \geq v'_k/c_i)~\forall\tau_k \in \mathcal{T}-C$
	\State $C \gets C \cup \tau_x$ 
	\If {$cost + c_x \leq W$}
		\State $cost \gets cost + c_x$
		\State $S \gets S \cup \tau_x$ 
	\EndIf	
\EndWhile
\end{algorithmic}
\end{algorithm}
\begin{algorithm}[t]
 \caption{Dynamic Greedy}
 \label{dynamic}
\begin{algorithmic}[1]
\State $\mathcal{T}$: list of all sources $\tau_i$
\State $C$: list of covered sources
\State $S$: set of selected sources
\State $N_i$: set of neighbors of source $\tau_i$
\State $v_i$: importance value of source $\tau_i$
\State $c_i$: cost of source $\tau_i$ 
\State $e_{ij}$: prediction edge (using source $\tau_i$ to predict source $\tau_j$)
\\
\State $cost \gets 0$
\While {$|\mathcal{T}| \neq |C|$}
    
	\For {each $ \tau_i \in \mathcal{T}-S $ }
		\State $v'_i \gets (1-e_{xi})v_i$  such that  $e_{xi}= \max\ (e_{ji})\ \forall \tau_j \in (S \cap N_i)\ or\ 0$
	\EndFor
	
    \For {each $ \tau_i \in \mathcal{T}-S $ }
   		 \State $ v'_x \gets v'_x + v_i\cdot \max (e_{xi}-e_{yi}, 0)$~such that 
   		 \State ~~~~~~~~$e_{xi} = \max~_{\tau_j \in (\mathcal{T} - S)} (e_{ji})~or~0,~\forall \tau_j \in N_i$ 
   		 \State ~~~~~~~~and $e_{yi} = \max~_{\tau_j \in S} (e_{ji})~or~0,~\forall \tau_j \in N_i$
	\EndFor

	 \State $x \gets j$ such that $(v'_j/c_i \geq v'_k/c_i)~\forall\tau_k \in \mathcal{T}-C$
	 \State $C \gets C \cup \tau_x$ 
    \If {$ cost + c_x \leq W $}
        \State $ S \gets S \cup \tau_x $ 
        \State $cost \gets cost + c_x$
    \EndIf    
\EndWhile
\end{algorithmic}
\end{algorithm}

\section{Evaluation}
\label{sec:evaluation}

In this section, we first evaluate the hybrid approach presented earlier versus the baselines described in Section~\ref{sec:statement} (i.e., Random, LastKnownState, BestProxy, Majority, and ARIMA) where we have no control over selection of POIs whose state is known. For ground truth, we use the aforementioned data set, featuring the daily status of gas stations, pharmacies, and food stores in the aftermath of Hurricane Sandy. As before, we opt to predict the status of these POIs on November 3rd and 8th, as examples of a day or relative calm and a day of significant change. We do so by varying the fraction of POIs whose state is revealed to the predictor on a given day, and attemtping to predict the rest using each of the compared approaches.

\begin{figure*}[!p!t]
\centering
\begin{minipage}{\textwidth}
\centering
\subfigure[][Error rate on Nov 3rd.]{
\includegraphics[width=0.48\textwidth]{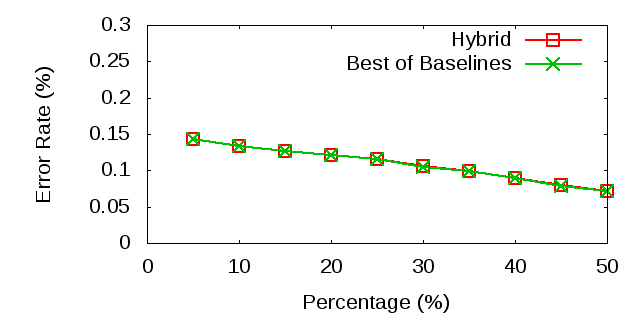}
\label{fig:gas4H}
}
\subfigure[][Error rate on Nov 8th.]{
\includegraphics[width=0.48\textwidth]{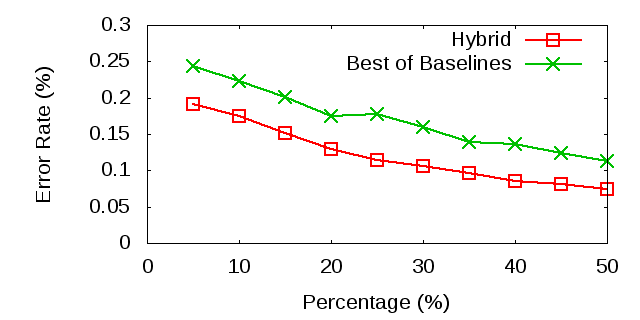}
\label{fig:gas9H}
}
\caption{Predicting gas availability after Sandy}
\label{fig:accuracy_hybrid_gas}
\end{minipage}
\begin{minipage}{\textwidth}
\centering
\subfigure[][Error rate on Nov 3rd]{
\includegraphics[width=0.48\textwidth]{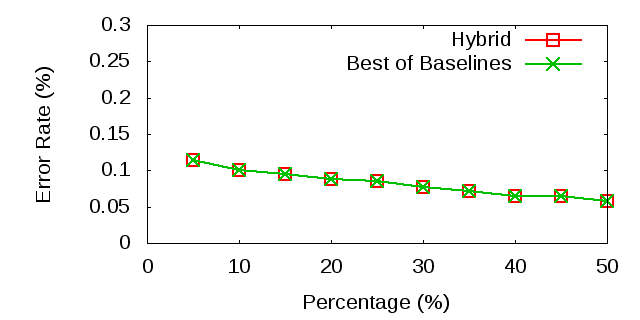}
\label{fig:food4H}
}
\subfigure[][Error rate on Nov 8th]{
\includegraphics[width=0.48\textwidth]{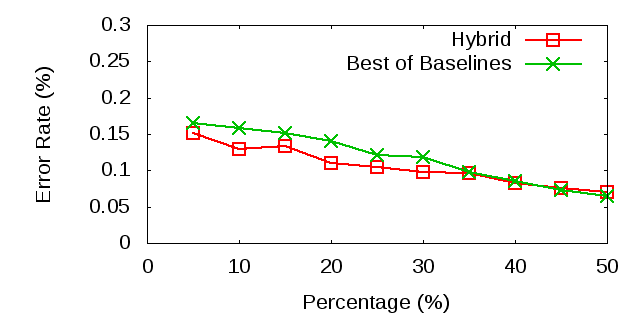}
\label{fig:food9H}
}
\caption{Predicting food availability after Sandy}
\label{fig:accuracy_hybrid_food}
\end{minipage}
\begin{minipage}{\textwidth}
\centering
\subfigure[][Error rate on Nov 3rd]{
\includegraphics[width=0.48\textwidth]{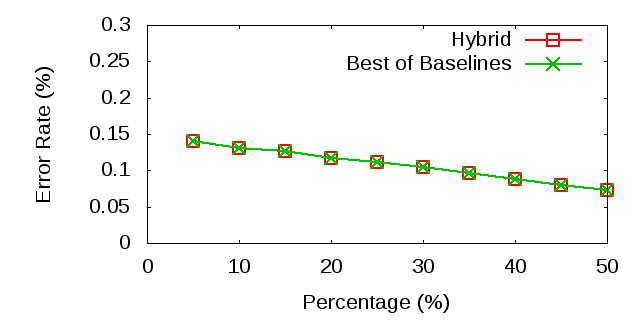}
\label{fig:pharma4H}
}
\subfigure[][Error rate on Nov 8th]{
\includegraphics[width=0.48\textwidth]{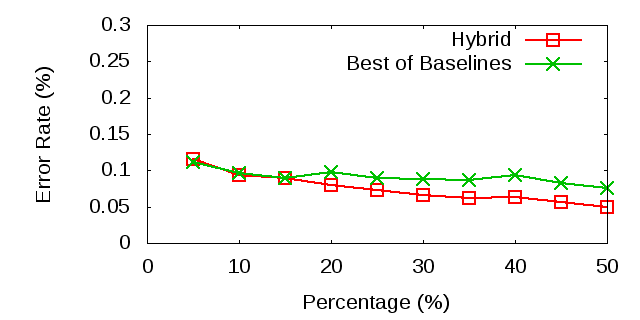}
\label{fig:pharma9H}
}
\caption{Predicting pharmacy availability after Sandy}
\label{fig:accuracy_hybrid_pharmacy}
\end{minipage}
\end{figure*}

Figures~\ref{fig:accuracy_hybrid_gas}-a and~\ref{fig:accuracy_hybrid_gas}-b illustrate the accuracy of prediction of gas availability on Nov 3rd and 8th, respectively. 
The horizontal axis shows the percentage of POIs whose state is known on the given day. As before, each point is the average of 50 experiments featuring different random selections of stations whose status is known. On the vertical axis, two curves are compared. One is the hybrid extrapolation algorithm developed in this paper. The second is the {\em best\/} of the predictions of the five baselines described in 
Section~\ref{sec:statement}. It can be seen that the new algorithm consistently matches or outperforms the best of all others.

Specifically, on November 3rd, the hybrid approach matches the best baseline. This is because it recognizes that change is small, and opts to use LastKnownState, which happens to be the best under the circumstances, as we have seen in Figure~\ref{fig:accuracy_gas}-a). On November 8th, it outperforms the best baseline, which tends to be BestProxy as we have seen in Figure~\ref{fig:accuracy_gas}-b. This is because of the new definition of correlation that it uses, which focuses only on changes, per {\em Insight \#2\/} discussed earlier.

Figures~\ref{fig:accuracy_hybrid_food}-a and~\ref{fig:accuracy_hybrid_food}-b repeat the experiment on the food data set. They illustrate the accuracy of prediction of food availability on Nov 3rd and 8th, respectively. A similar trend is seen, where the hybrid matches the best baseline on November 3rd and outperforms the best baseline on November 8th. 
Figures \ref{fig:accuracy_hybrid_pharmacy}-a 
and~\ref{fig:accuracy_hybrid_pharmacy}-b illustrate the same for pharmacies. Further experiments (not shown) demonstrated that the results are largely insensitive to the choice of threshold, $T$. The superior results presented above can therefore be robustly achieved. 

The experimental results presented in this section show that the hybrid approach is as good as or better than the best of all compared algorithms on both November 3rd and November 8th. These two days were selected because of their representative nature, as they exemplified days of calm and days of change, respectively. 

\begin{figure}[p]
\begin{center}
\begin{tabular}{c}
  \includegraphics[totalheight=0.3\textheight]{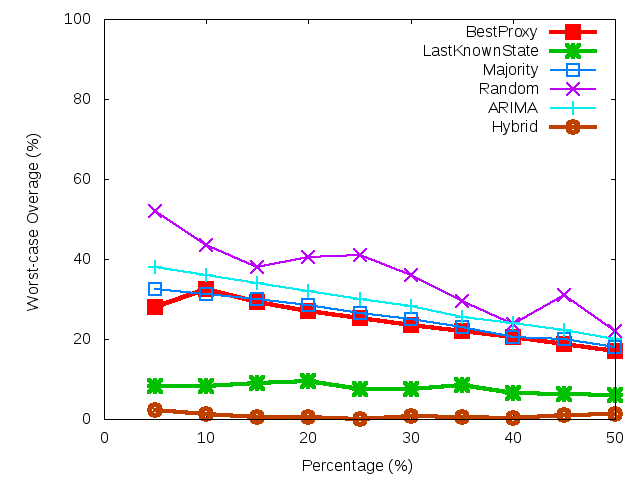}%
\\
a) Worst-case overage in gas availability prediction error.
\\
  \includegraphics[totalheight=0.3\textheight]{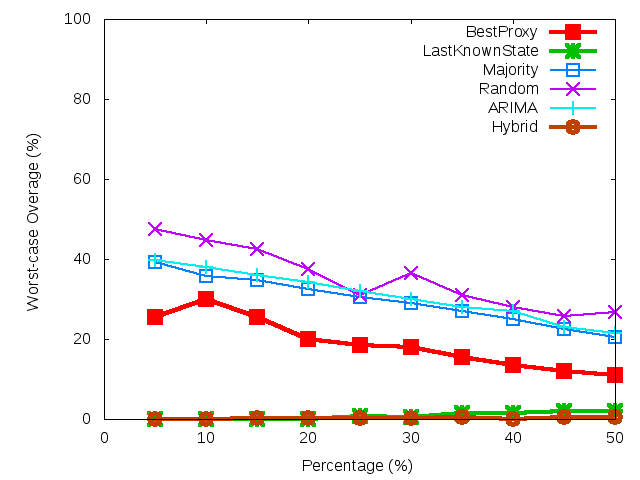}%
\\
b) Worst-case overage in food availability prediction error.
\\
\includegraphics[totalheight=0.3\textheight]{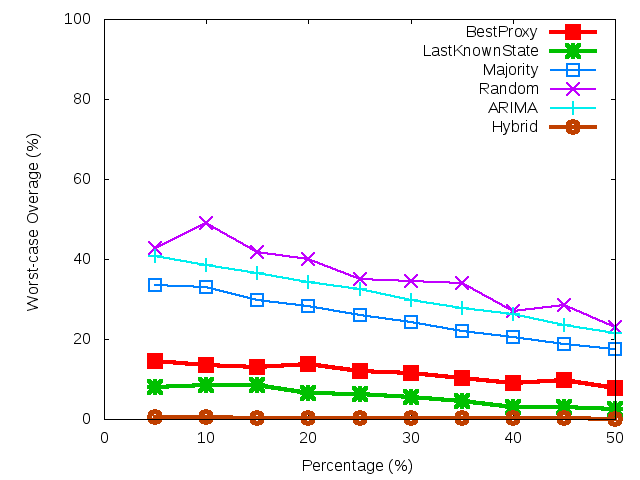}%
\\
c) Worst-case overage in pharmacy availability prediction error.
\end{tabular}
\caption{Worst-case prediction error overage of individual solutions}
\label{fig:best}
\end{center}
\end{figure}

To show that the above results hold true for other days as well, we compute the {\em worst case\/} overage amount by which the prediction error of the hybrid approach, as well as the prediction error of each of the five individual baselines, exceeds the best of the five baselines. 
Hence, an algorithm that behaves as the best of the baselines under all circumstances will have a worst-case overage of zero. Algorithms that are not consistently the best will have a higher worst-case overage. The results are shown in Figure~\ref{fig:best}, where Figure~\ref{fig:best}-a, Figure~\ref{fig:best}-b, and Figure~\ref{fig:best}-c, are for the case of gas, food, and pharmacy availability prediction, respectively. 

In Figure~\ref{fig:best}, the worst-case overage, for each algorithm, is computed by finding the maximum error overage computed over 10 days of the recovery phase (from November 3rd through November 12th). For statistical significance, the performance of each heuristic on each day is first averaged over 50 experiments before the overage is calculated. Consistently with other figures, the horizontal axis shows the percentage of PoIs whose status is known. It is seen that the new Hybrid algorithm has a worst-case overage that is roughly zero. In other words, {\em it never does worse than the best solution over all days under consideration\/}. 

\begin{figure}[p]
\begin{center}
\begin{tabular}{c}
\includegraphics[totalheight=0.3\textheight, trim = 50 225 50 225, clip = true]{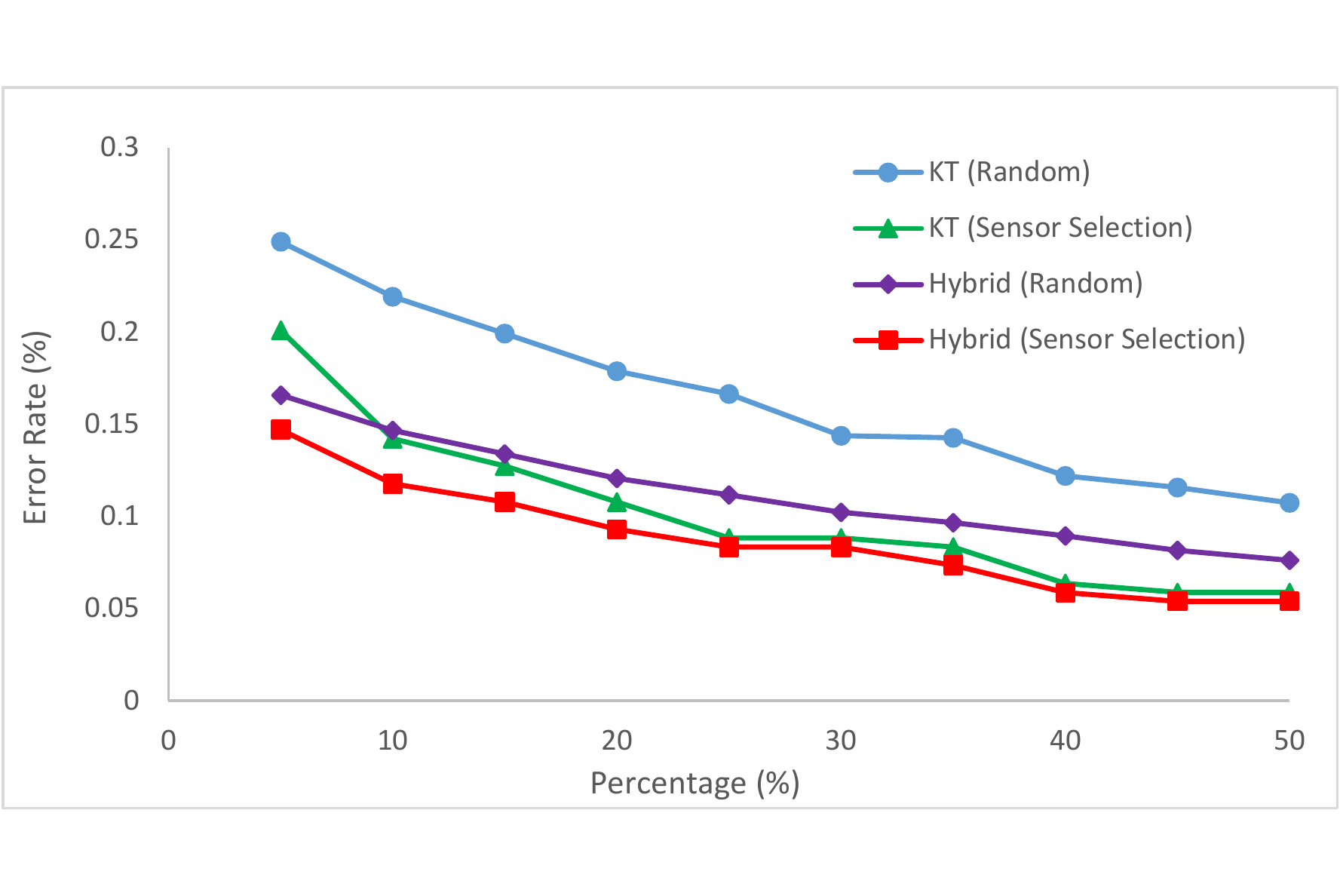}\\
a) prediction errors for gas availability on Nov 8th.\\
\includegraphics[totalheight=0.3\textheight, trim = 50 225 50 225, clip = true]{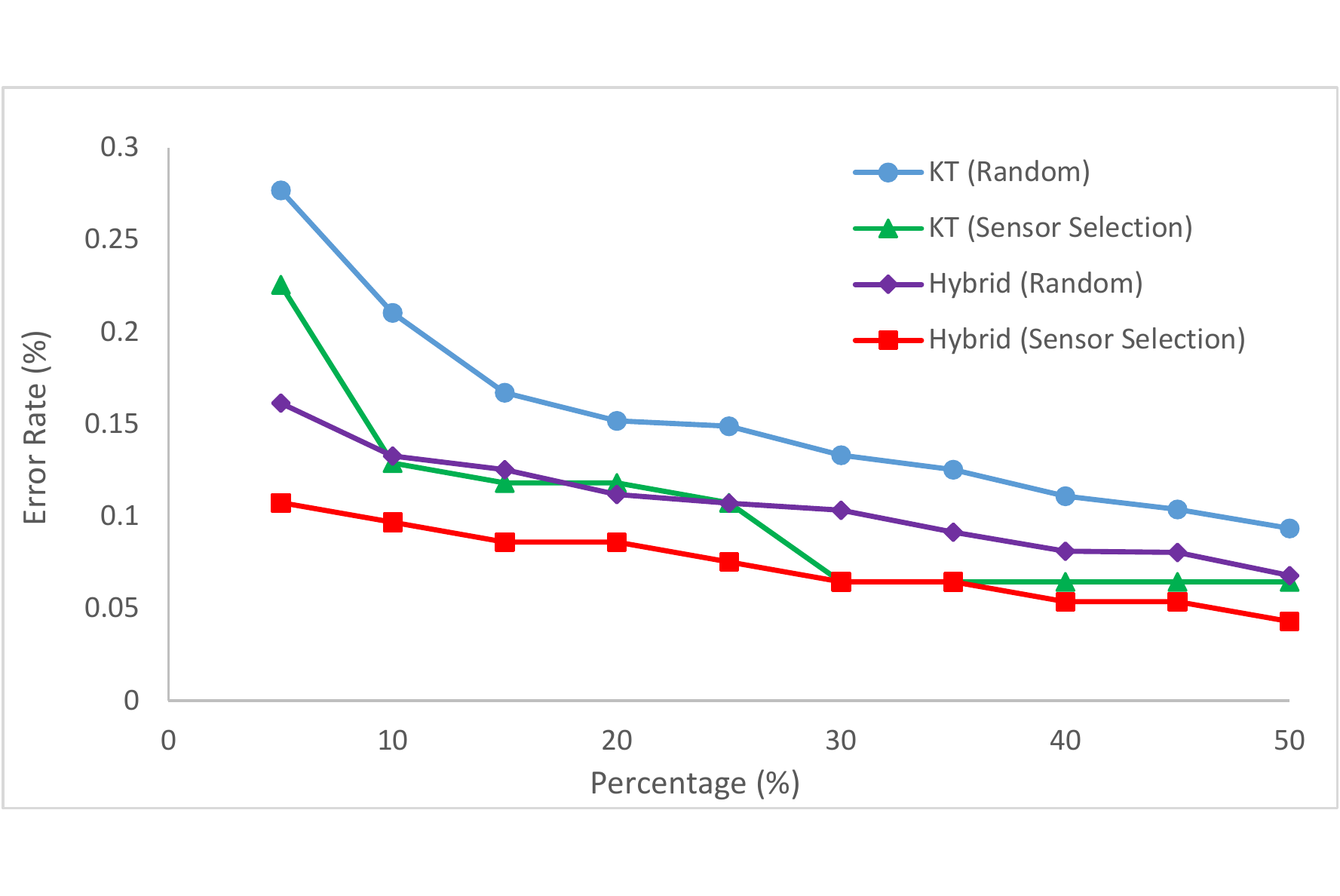}\\
b) prediction errors for food availability on Nov 8th.\\
\includegraphics[totalheight=0.3\textheight, trim = 50 225 50 225, clip = true]{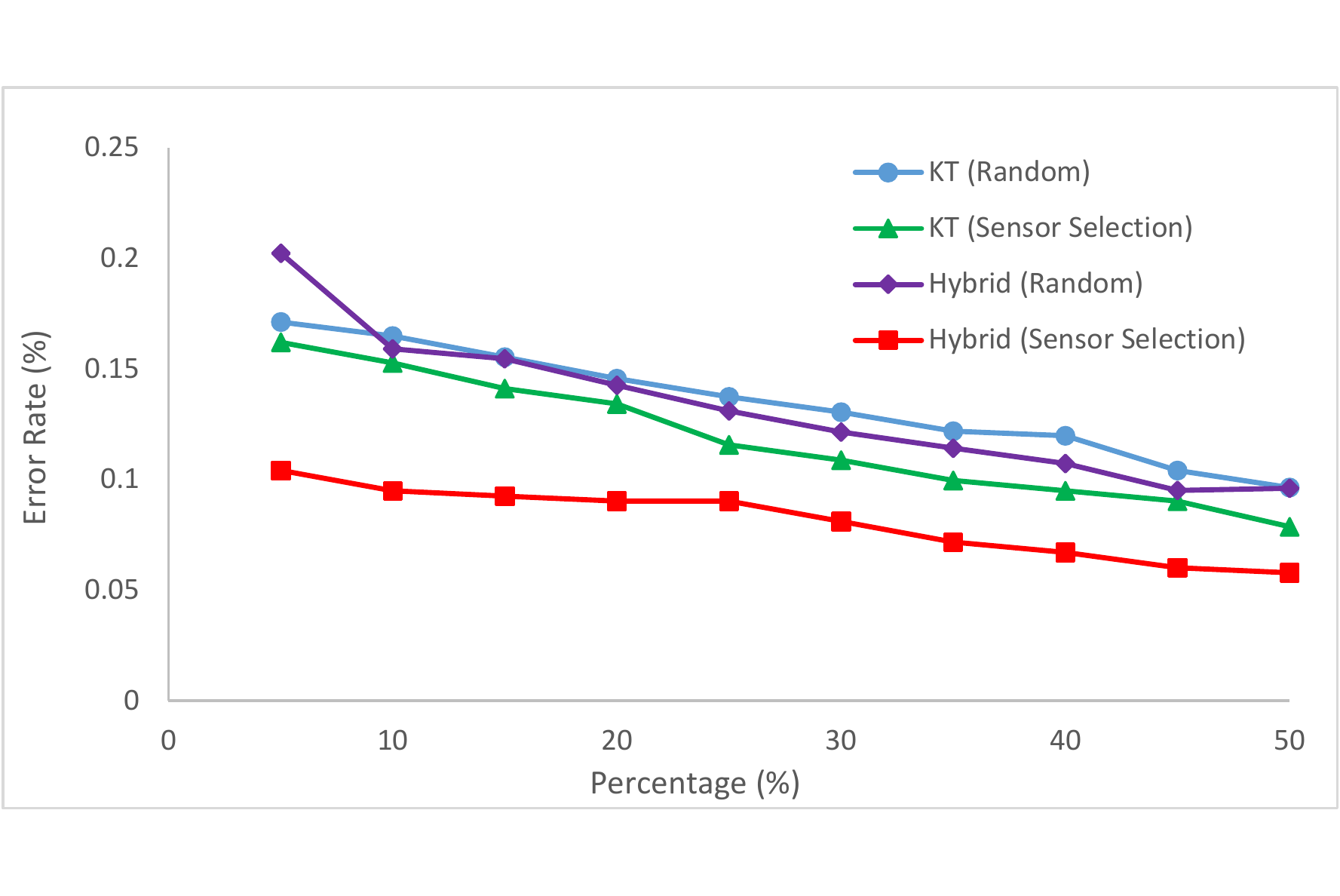}\\
c) prediction errors for pharmacy availability on Nov 8th.\\
\end{tabular}
\caption{A comparison of prediction errors for gas, food, and pharmacy availability on Nov 8th, using \textit{Static} greedy algorithm, measured for both KT and Hybrid correlations.}
\label{static-alg}
\end{center}
\end{figure}

\begin{figure}[p]
\begin{center}
\begin{tabular}{c}
\includegraphics[totalheight=0.3\textheight, trim = 50 225 50 225, clip = true]{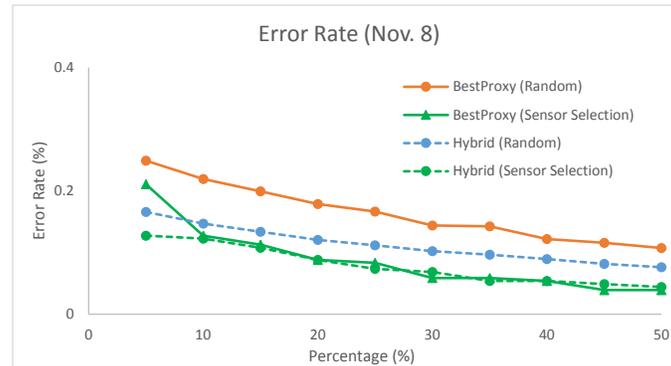}\\
a) prediction errors for gas availability on Nov 8th.\\
\includegraphics[totalheight=0.3\textheight, trim = 50 225 50 225, clip = true]{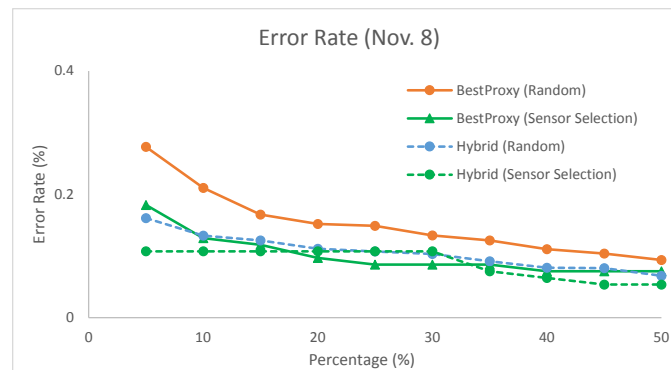}\\
b) prediction errors for food availability on Nov 8th.\\
\includegraphics[totalheight=0.3\textheight, trim = 50 225 50 225, clip = true]{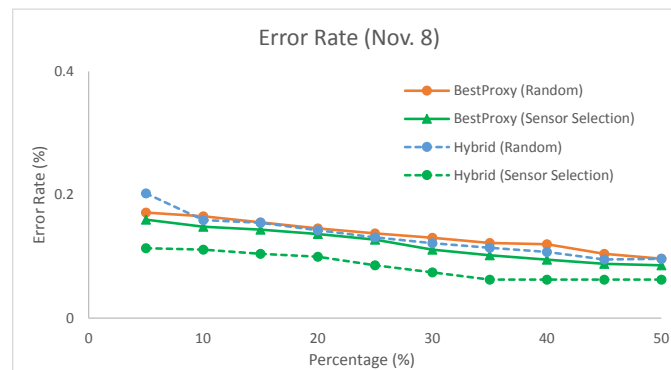}\\
c) prediction errors for pharmacy availability on Nov 8th.\\
\end{tabular}
\caption{A comparison of prediction errors for gas, food, and pharmacy availability on Nov 8th, using \textit{Dynamic} greedy algorithm, measured for both Best Proxy and Hybrid correlations.}
\label{dynamic-alg}
\end{center}
\end{figure}

\begin{figure}[p]
\begin{center}
\begin{tabular}{c}
\includegraphics[totalheight=0.3\textheight, trim = 50 225 50 225, clip = true]{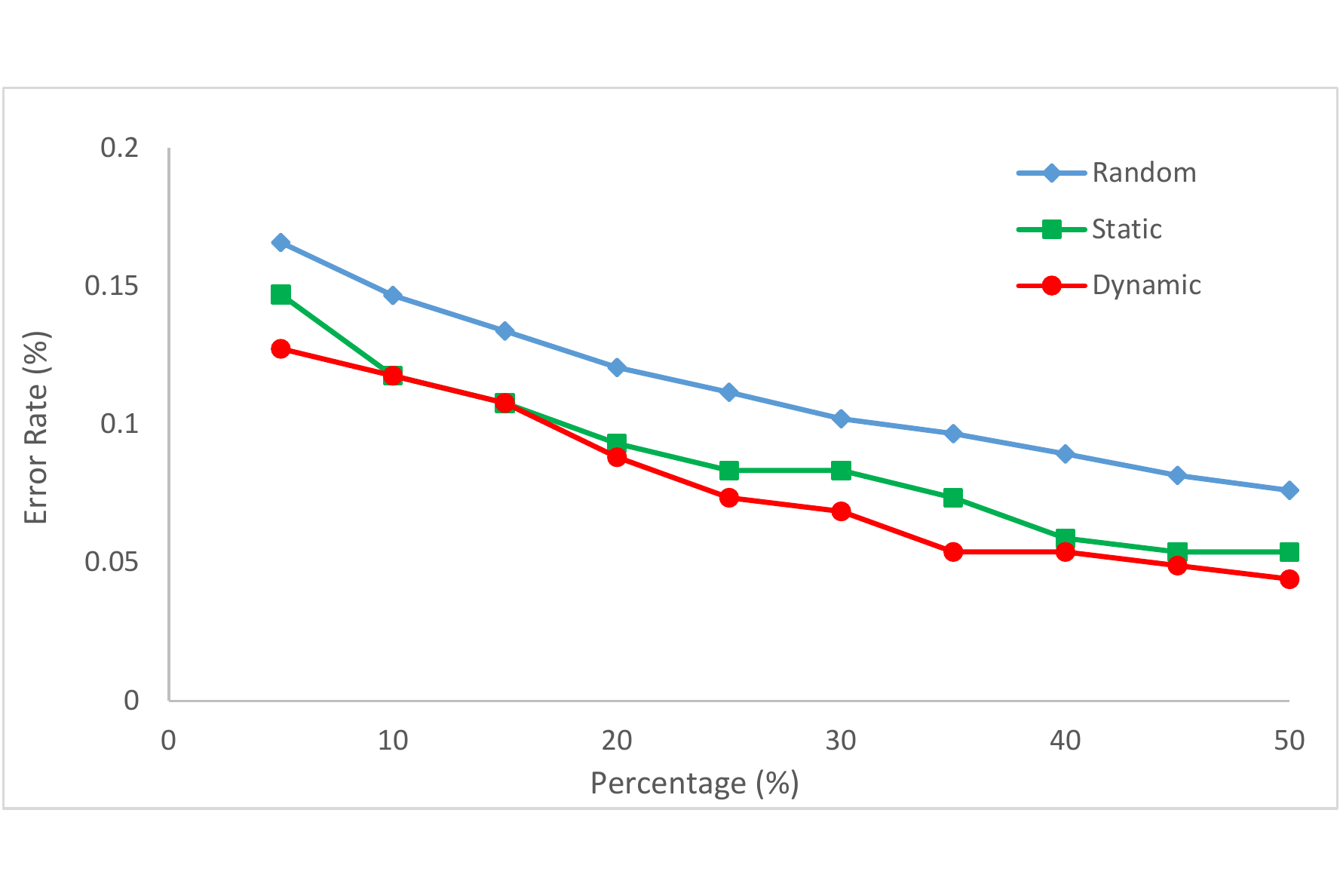}\\
a) prediction errors for gas availability on Nov 8th.\\
\includegraphics[totalheight=0.3\textheight, trim = 50 225 50 225, clip = true]{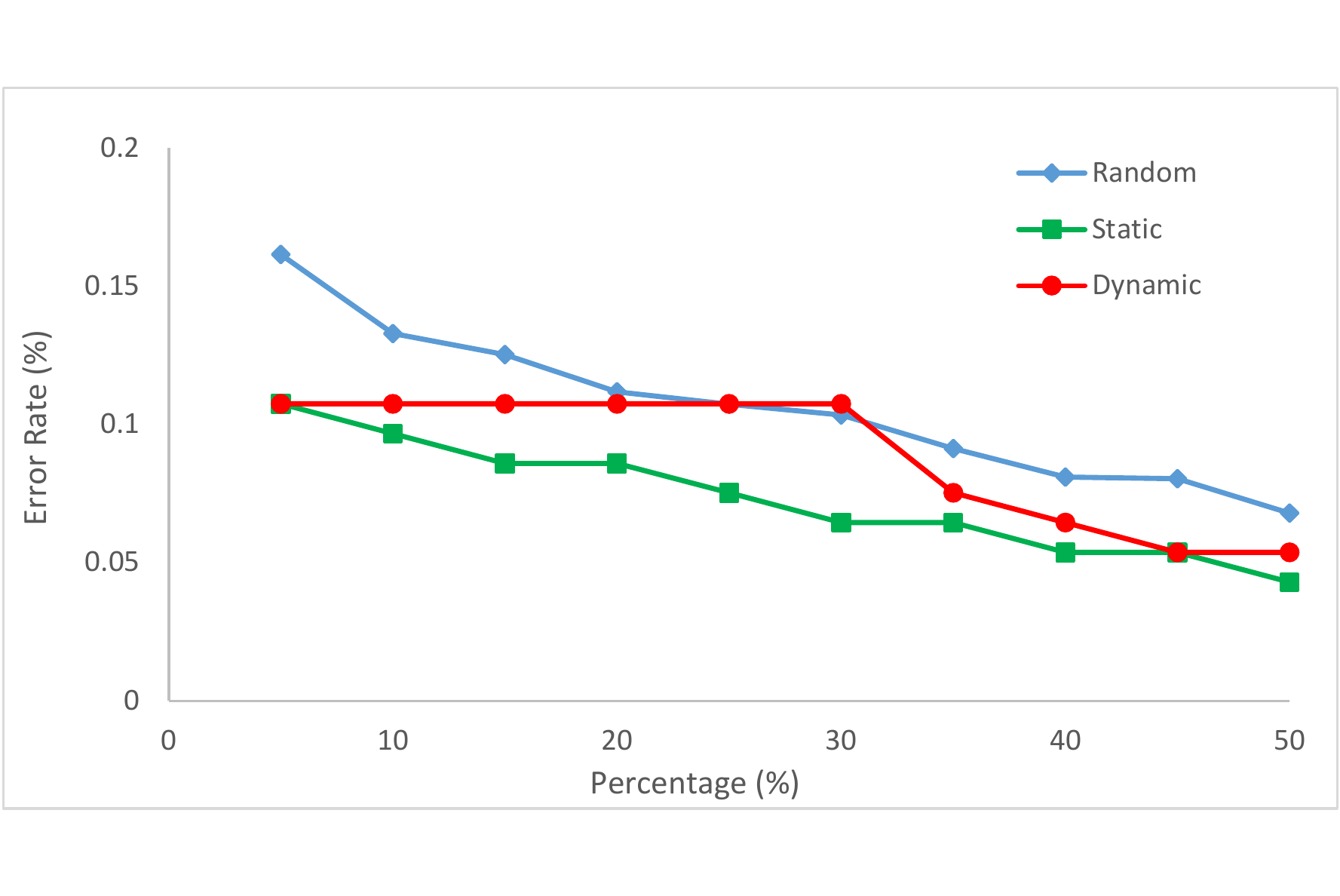}\\
b) prediction errors for food availability on Nov 8th.\\
\includegraphics[totalheight=0.3\textheight, trim = 50 225 50 225, clip = true]{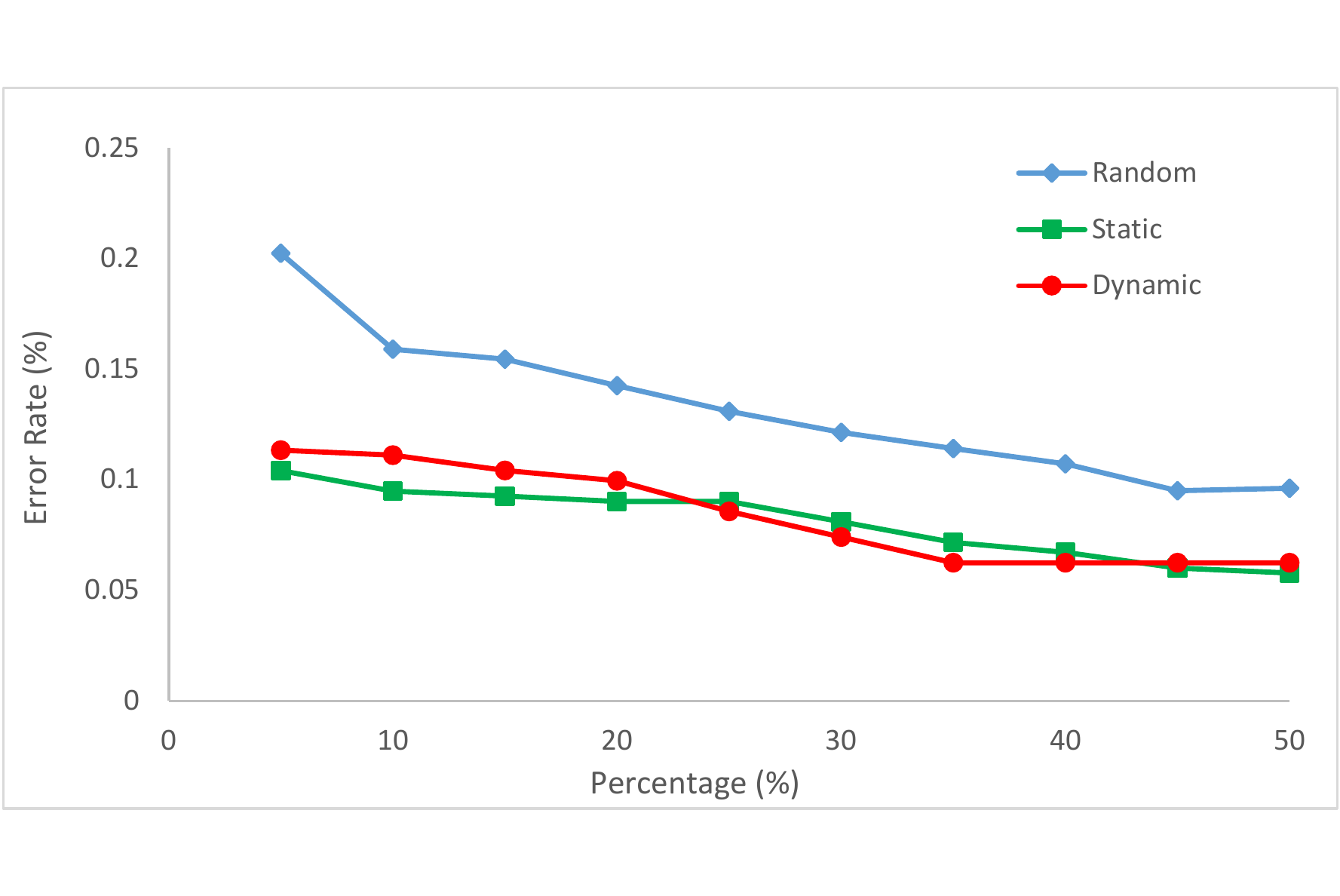}\\
c) prediction errors for pharmacy availability on Nov 8th.\\
\end{tabular}
\caption{A comparison of prediction errors for gas, food, and pharmacy availability on Nov 8th, for Static and Dynamic greedy algorithms against Random baseline using Hybrid correlation.}
\label{pdf2}
\end{center}
\end{figure}

The figure shows that the overage of other baselines is higher. 
Their relative prediction (in) accuracy follows roughly the same order in the three data sets. Specifically, LastKnownState is generally the next best algorithm to ours. In the aftermath of disasters, failures take long to fix, so the state changes gradually, making LastKnownState a good predictor most of the time. Errors occur when aftershocks hit or major repairs are made, and are related to the size of such perturbations.  BestProxy comes next. Its accuracy depends on how spatially well-correlated the POI states are. No significant difference is seen between its accuracy in gas and food availability prediction, but pharmacy prediction is better. This can be attributed to the size of the pharmacy data set, shown on the horisontal axis in Figure~\ref{fig:fig1_pharma}. Namely, the number of pharmacies is the largest. Hence, the odds of finding a good proxy are better than with the other data sets. Majority comes next after BestProxy. In scenarios where restoration is quicker, PoIs converge to the majority state faster, and the predictor becomes more accurate. Comparing Figure~\ref{fig:fig1_gas},~\ref{fig:fig1_food}, and~\ref{fig:fig1_pharma}, we can see that pharmacies and gas are restored the fastest, followed by food, which roughly corresponds to how well Majority works in the three cases. Finally, ARIMA and Random consistently do next-to-worst and worst, respectively, showing little variation acorss the data sets. This is because their worst-case behavior is random (for ARIMA, it occurs in the very early days), and hence not tightly related to the properties of input data.

In conclusion, Figure~\ref{fig:best} shows that while some prediction algorithms do best under some circumstances, no baseline does consistently well under all circumstances.

To evaluate our proposed source selection algorithms, we repeat the same experiments for the accuracy of prediction as in Figures \ref{fig:accuracy_hybrid_gas} to \ref{fig:accuracy_hybrid_pharmacy} for availability of resources on Nov 8th. Figures \ref{static-alg}(a) to \ref{static-alg}(c) illustrate how prediction error using our proposed \textit{Static} source selection algorithm compares with that of the baseline random selection for availability of gas, food, and pharmacy, respectively. As the figures suggest, our Static greedy source selection algorithm significantly improves the overall system predictability and minimizes the prediction errors, and shows that our proposed Hybrid correlation metric is performing better than the KT-based metric. 
In a similar way, we also evaluate the performance of Dynamic source selection algorithm. Figures \ref{dynamic-alg}(a) to \ref{dynamic-alg}(c) show the prediction error using our proposed \textit{Dynamic} source selection algorithm compared with that of the baseline random selection for availability of gas, food, and pharmacy. As can be see in the figures, the Dynamic source selection algorithm outperforms the baseline algorithm, with BestProxy-based random selection scheme and Hyrbid-based source selection scheme having the highest and lowest prediction errors, respectively.

We also compare the Static and Dynamic greedy algorithms against the Random baseline scheme in which none of our source selection algorithms is applied. Figures \ref{pdf2}(a) to \ref{pdf2}(c) show the evaluation results. As the figures suggest, our proposed source selection algorithms provide extended improvement on the baseline algorithm by adaptively selecting the best node candidates given the percentage of retrieved sources. As can be noticed in Figure \ref{pdf2}(b) and Figure \ref{pdf2}(c), our proposed Static source selection algorithm slightly outperforms the Dynamic algorithm for some cases. The interesting point here is that depending on the case, a combination of both Static or Dynamic source selection must be used in order to achieve lowest prediction error. This fact opens a new area of research to further seek a \textit{Hybrid Source Selection} scheme in which a combination of Static and Dynamic selection algorithms is used to maximize the overall prediction. We leave this interesting concluding part as the future work.


In summary, the paper contributed new algorithms that: (a) adapt intelligently between time-based extrapolation and spatial extrapolation, matching or outperforming the best baseline solutions, and (b) adaptively select the best sources to retrieve for prediction of state of other nodes using our proposed source selection algorithms.

\section{Related Work}
\label{sec:soa}

Our work extends a large body of research on sensor network that focused on monitoring and disaster alerts. For example, Werner-Allen et al. deployed three wireless sensor networks on active volcanoes~\cite{Werner-Allen:OSDI}. The initial deployment was a small proof of concept system that monitored acoustic signals from the Tungurahua volcano, in Ecuador. The second deployment was to measure seismic signals at the Reventador volcano, in Ecuador. The third deployment was at Tungurahua in August, featuring a new data collection system. Li et al. deployed a sensor network for monitoring and alerts in a coal mine~\cite{Li:TOSN}. Liu et al. present an automatic and reliable sensor network for firefighter applications~\cite{Liu:MobiSys}, which allows a firefighter to carry a small dispenser filled with sensor nodes and deploy them one-by-one in a manner that guarantees reliable communication. The SensorFly project~\cite{Purohit:IPSN} develops a sensor cloud, which consists of many low cost and individually limited mobile sensing devices that only when functioning together can produce an intelligent cloud, in disaster situations such as an
earthquake and fire. This paper is different from the above work in leveraging a participatory sensing framework, and considering first responders and volunteers as front-end sensors for data collection.

More importantly, our work focuses on a new problem in participatory sensing. Namely, the problem of automatically filling in the ``blind spots''
in reported observations. Past research on participatory sensing
describes how to aggregate and clean-up collected data. A survey on analytic challenges in the field recently appeared~\cite{charu:13}. For instance, CenWits~\cite{Huang:SenSys} proposes a participatory sensor network to rescue hikers in emergency situations. BikeNet~\cite{Eisenman:SenSys} presents a bikers sensor network for sharing cycling related data and mapping
the cyclist experience. The Nericell project~\cite{Mohan:SenSys} presents a
system that performs rich sensing using smartphones that users carry
with them in normal course, to monitor road and traffic conditions.
The GreenGPS system~\cite{Ganti:MobiSys} provides a service that computes
fuel-efficient routes for vehicles between arbitrary end-points, by
exploiting vehicular sensor measurements available through the On
Board Diagnostic (OBD-II) interface of the car and GPS sensors on
smart phones. SignalGuru~\cite{Balan:MobiSys} is a software service that relies
solely on a collection of mobile phones to detect and predict the
traffic signal schedule, producing a Green Light Optimal Speed
Advisory (GLOSA). This paper complements that past work by looking
at the important problem of how to fill in the data gaps. This
unique challenge comes from the timeliness constraints in disaster
response applications. In the absence of urgency, one can eventually
fill in the data gaps by sending (or waiting for) more observers.
Hence, there is less need to ``guess''  them. However, in disaster
recovery scenarios, there is no time to wait, so the service
provider needs to fill in the gaps immediately as best one can.

Thanks to the fast development of smartphones and social networks, participatory sensing receives more attention in disaster response applications in recent years. People share their information about the disaster region to social networks and special-purpose services, to help each other beat the disaster together. For instance, popular social networks such as Facebook~\cite{facebook:web} and Twitter~\cite{twitter:web}, played an important role after natural disasters such as Japan Tsunami in 2011~\cite{Tsunami:web} and US Hurricane Sandy in 2012~\cite{Sandy:web}. Many service providers, some notable names including Waze~\cite{Waze:web} and GasBuddy~\cite{gasbuddy:web}, set up special-purposes services to allow individuals to participate and report the availability of various resources (e.g., gas stations) after Sandy via the web or smartphones. However, due to the opportunistic nature of participatory sensing, there are typically ``blind points'' in the obtained PoI map at any given time point. Our work takes advantage of these services, aiming to complete the estimation of missing world state.

Finally, our system design is related to state of the art sensor selection algorithms that are paired with inference approaches for missing or incomplete data. For example, Aggarwal {\em et al.\/} formulate the problem of sensor selection, when redundancy relationships between sensors can be expressed through an information network by using external linkage information. They present methods for sensor selection by using regression models to estimate predictability and redundancy~\cite{Aggarwal:DCOSS}. Aggarwal {\em et al.\/} also presented algorithms for data-driven selection of sensor streams to collect the data passively at low sampling rates in order to detect any changing trends in the underlying data. However, the goal there was to minimize the power consumption in reducing the data collected~\cite{Aggarwal:kdd}. Our new contributions in this paper extends their work to both static and dynamic sensor selection with the aim of reducing redundancies. Similarly, PhotoNet~\cite{Uddin:RTSS} provides a picture-collection service for disaster response applications that maximizes situation-awareness. Kobayashi et al. propose a sensor selection method with fuzzy inference for sensor fusion in robot applications~\cite{Kobayashi:Fuzzy}. However, this existing work assumes that correlations between data items are known in advance. These correlations are the basis for sensor selection. Also, they assume a stationary process. 

Biswas et al. proposed a Bayesian inference approach and applied it on a simulated problem of determining whether a friendly agent is surrounded by enemy agents~\cite{Biswas:IPSN}. However, their approach does not work for binary PoI information due to the logistic regression overflow problem. 

To the best of our knowledge, no previous work has been applied to real-world disaster response scenarios where inference algorithms were investigated that (i) specifically address the bimodal nature of damage propagation and that (ii) require very little training data. Our paper fills in this gap by analyzing the example of New York City gas crisis in the aftermath of Hurricane Sandy via real data traces.

\section{Conclusions}
\label{sec:conclusion}

We presented the design, implementation, and evaluation of inference-based algorithms for data extrapolation and prediction in participatory sensing systems for disaster response applications. They are shown to be capable of accurately predicting the status of POI sites, when collected data is incomplete. We proposed a hybrid algorithm that exploits correlations among state changes in POI sites and changes adaptively between temporal and spatial extrapolation. We also described two classes of source selection algorithms that further enhance the state inference of unknown POIs. Our experimental results via a real-world disaster response application demonstrate that our algorithms are consistently the best of all compared in terms of prediction accuracy, whereas others may suffer non-trivial degradation. The new algorithms are currently being adapted to more complex prediction tasks (e.g., non-binary variables) and evaluated on new data sets.

\newpage
\bibliographystyle{ACM-Reference-Format-Journals}
\bibliography{ref}

\end{document}